\documentclass[showpacs,preprintnumbers,preprint,nofootinbib,
superscriptaddress,showkeys]{revtex4}

\def\versioninfo{hen-23.tex}

\usepackage{amssymb}
\usepackage{amsmath}
\usepackage{graphicx}
\usepackage{dcolumn}
\usepackage{bm}
\usepackage{epsf}

\def\be{\begin{eqnarray}}
\def\ee{\end{eqnarray}}
\def\bea{\begin{eqnarray}}
\def\eea{\end{eqnarray}}
\def\bm{\boldsymbol}

\def\vR{{\bm R}}

\def\vr{{\bm r}}

\def\vk{{\bm k}}
\def\vp{{\bm p}}
\def\vbp{{\overline {\bm p}}}
\def\vx{{\bm x}}

\def\vy{{\bm y}}

\def\vs{{\bm \sigma}}
\def\vmu{{\bm \mu}}

\def\hatr{{\hat \vr}}
\newcommand{\no}{\nonumber \\}
\newcommand{\etal}{{\it et al.}}

\def\H#1{{}^{#1}\mbox{H}}
\def\He#1{{}^{#1}\mbox{He}}
\def\nlo#1{\mbox{N$^{#1}$LO}}

\def\dR{{\hat d}^R}
\def\TminusS{{{\hat T}_S^{(-)}}}
\def\TminusT{{{\hat T}_T^{(-)}}}
\def\TplusS{{{\hat T}_S^{(+)}}}
\def\TplusT{{{\hat T}_T^{(+)}}}
\def\TtimesS{{{\hat T}_S^{(\times)}}}
\def\TtimesT{{{\hat T}_T^{(\times)}}}
\def\fm{{\mbox{fm}}}
\def\calM{\mathcal{M}}

\begin{document}

\preprint{\versioninfo}

\title{Heavy-baryon chiral perturbation theory approach to 
thermal neutron capture on ${}^{3}\mbox{He}$}
\author{Rimantas Lazauskas}
\email{rimantas.lazauskas@ires.in2p3.fr}
\affiliation{IPHC, IN2P3-CNRS/Universit\'e Louis Pasteur B.P. 28,
F-67037 Strasbourg Cedex 2, France}
\author{Young-Ho Song}
\email{yhsong@phy.duke.edu, song25@mailbox.sc.edu}
\affiliation{Department of Physics, Duke University, Durham, NC7708, USA}
\affiliation{Department of Physics and Astronomy,
University of South Carolina, Columbia, SC 29208, USA}
\author{Tae-Sun Park}
\email{tspark@kias.re.kr}
\affiliation{Department of Physics and BAERI, Sungkyunkwan University, Suwon 440-746,
Korea}
\affiliation{Department of Physics and Astronomy,
University of South Carolina, Columbia, SC 29208, USA}

\date{\today }

\begin{abstract}
The cross section for radiative thermal neutron capture on
${}^{3}\mbox{He}$ ($\He3 +n \to \He4 +\gamma$;
known as the $hen$ reaction)
is calculated based
on heavy-baryon chiral perturbation theory.
The relevant M1 operators are derived up to
next-to-next-to-next-to-leading order (N${}^3$LO).
The initial and final nuclear wave functions are obtained
from the rigorous Faddeev-Yakubovski equations
for five sets of realistic nuclear interactions.
Up to N${}^3$LO, the M1 operators
contain two low-energy constants,
which appear as the coefficients
of non-derivative two-nucleon contact terms.
After determining these two constants
using the experimental values of the magnetic moments of the triton
and ${}^3\mbox{He}$,
we carry out a  parameter-free calculation
of the $hen$ cross section.
The results are in good agreement with the data.
\end{abstract}

\maketitle

\renewcommand{\thefootnote}{\#\arabic{footnote}} \setcounter{footnote}{0}

\section{Introduction}\label{sec:intro}

The radiative capture of a slow neutron on $\He3$ ($\He3 + n \to \He4 + \gamma$),
or the $hen$ process,
is an example of rare situations where
the contributions of the single-nucleon (1B) currents
are strongly suppressed owing to the so-called pseudo-orthogonality,
which refers to the fact that
the major components of the initial and final states belong to different
representations of the spatial symmetry group and
hence cannot be connected
by the $r$-independent leading 1B operators.
To be more specific, the $hen$ reaction proceeds
from a $J^\pi=1^+$ $n$+$\He3$ state whose
dominant component belongs to a $[31]$ representation
of the spatial permutation group,
while the final $\alpha$-particle state belongs to
a $[4]$ representation, and these two representations
cannot be connected by the leading Gamow-Teller operator,
$\vec{\tau}\vec{\sigma}$.
This suppression is so drastic that
the meson-exchange-current (MEC) ``corrections" become
comparable to the 1B contributions.
Furthermore, it turns out that the MEC and 1B terms in this case
come with opposite signs,
leading to a further drastic suppression
of the $hen$ cross section.

The $hen$ process near threshold
is governed by the M1 operators
(as both the initial and the final states
are dominated by the $S$ waves at low energy),
and the MEC contributions to them consist of
the well-known one-pion-exchange part
and the short-range part.
Note that the latter is not constrained
by the symmetries of QCD, and that,
because of the above-mentioned suppression and
the cancellation mechanisms,
the short-range contributions are crucially important
even for a rough estimation of the cross section.
Furthermore, the strong suppression of
the 1B matrix elements (MEs) implies that
their values are
sensitive to the details
of the wave functions.
Therefore, for a precise estimation of the $hen$ cross section,
it is imperative to have:
(i) a reliable method for deriving the relevant MEC operators
with a good control of short-range physics,
and (ii) the accurate wave functions for the initial and final
nuclear states.
These requirements make the quantitative estimation of the
$hen$ cross section highly nontrivial.

At the same time,
the strong enhancement of the relative importance of MEC in
the $hen$ process makes it a valuable testing ground for
the roles of MEC in light nuclei.
Apart from this point, which is important on its own right,
a careful study of the $hen$ process is also of great
significance
in connection
with the so-called $hep$ processes,
$\He3 + p \to \He4 + \nu_e + e^+$,
because $hep$ shares all the aforementioned features of $hen$:
the drastic suppression of the 1B contributions, strong cancellation
between the 1B and the 2B contributions, and the sensitivity of the
transition amplitude to the details of the nuclear wave functions.
The $hep$ process is one of the proton-burning reactions that take place
in the interior of the sun, and because $hep$ produces
the highest-energy solar neutrinos,
it has attracted much attention in the study of
the solar neutrinos
(see Refs. \cite{KP,INT09} for a recent review)
and motivated
a series of elaborated studies
~\cite{carlson-hep,lm-prl84,lm-prc63,Schiavilla:1992sb}.
Park {\it et al.}~\cite{hep}
developed an effective field theory (EFT) approach,
which has come to be known
as ``more-effective EFT" (MEEFT for short)~\cite{BR03}
and, with the use of MEEFT,
they calculated the $hep$ $S$-factor
with an estimated accuracy of about 15 \%.
In view of the fact that the previous theoretical estimations of the $hep$ $S$-factor
ranged over two orders of magnitude~\cite{hep},
this is a remarkable feat.
A direct test of this theoretical result, however, is not possible
because of the unfeasibility of the $hep$ cross section measurement.
Meanwhile, the threshold $hen$ cross section
has been measured with reasonable accuracy:
$\sigma_{exp}=(54\pm 6)~\mu b$ \cite{Wolfs} and
$\sigma_{exp}=(55\pm 3)~\mu b$ \cite{Wervelman}.
Given the close similarity between $hen$ and $hep$,
it is expected that comparison
between theory and experiment for the $hen$ case
offers valuable information on the validity of the theoretical framework
employed for the $hep$ calculation in Ref. \cite{hep}.
This gives an additional motivation
for carrying out a detailed study of $hen$.

Although the theoretical investigation of $hen$ has a long history,
the $hen$ cross section has never been explained in a satisfactory manner.
Towner and Khanna~\cite{Towner:1981hz}
and Wervelman \etal~\cite{Wervelman}
performed shell-model calculations
for schematic Hamiltonians and
obtained $\sigma=(14\sim 125)\ \mu b$ and $\sigma=(47\pm18)\ \mu b$,
respectively.
Much more elaborate calculations
with the use of  realistic Hamiltonians
have been performed by
Carlson \etal~\cite{Carlson:1990nh} and by
Schiavilla \etal~\cite{Schiavilla:1992sb},
who arrived at $\sigma \!=\!112~\mu b$ and
$\sigma\!=\!86~\mu b$, respectively.
These works are based on the conventional approach,
the so-called {\it standard nuclear physics approach} (SNPA for short),
which consists in the use of  phenomenological transition operators
and phenomenological wave functions.
SNPA has been enormously successful
in correlating and explaining a vast range of
electroweak nuclear transitions in nuclei
but, from a formal point of view,
it has an insufficiency that
it is not equipped with a systematic way of
reducing the uncertainty in the MEC operators.
The MEC operators in SNPA are constructed
based on the approach of Chemtob and Rho~\cite{cr71}.
Although this construction of the MEC operators
is guided by chiral symmetry and the Ward identities,
it is in general unavoidable to have ``model-dependent terms".

In this paper we report on a parameter-free MEEFT calculation for the
$hen$ cross section at threshold,
adopting essentially the same method
as used in \cite{hep}.\footnote{There has been an attempt
by Song et al.~\cite{Song:2003ja}
to carry out an MEEFT calculation of $hen$,
but an approximate treatment of the scattering wave function
in~\cite{Song:2003ja} limits its reliability.}
In MEEFT the transition operators are derived from the systematic expansion
of the heavy-baryon chiral perturbation theory (HBChPT),
and the nuclear MEs
are obtained by sandwiching these operators
between the wave functions generated
from a high-precision phenomenological nuclear potential.
Mismatch in the short-range part of the wave function
is overcome by the renormalization procedure
for the local operators, according to the premise of low-energy EFTs
(see below).
Thus MEEFT takes advantage of the systematic nature of EFTs
and the availability of state-of-the-art wave functions.
The mentioned ``formal" mismatch may be regarded as a weak point,
and also the accurate reproduction of the relevant
effective-range parameters (ERPs)
is not automatically
guaranteed in MEEFT; we come back to these points later.

In the present work we derive the M1 operators within HBChPT up to \nlo3.
These M1 operators turn out to contain two low-energy constants (LECs),
denoted $g_{4s}$ and $g_{4v}$,
which are the coefficients of two-nucleon contact-term operators.
These LECs can be fixed by requiring that the experimental values of
the magnetic moments of the triton and $\He3$,
$\mu(\H3)$ and $\mu(\He3)$,  be reproduced;
this is the same strategy as adopted in Refs.~\cite{slp1,slp2},
where the M1 properties of the $A$=$2$ and $A$=$3$ systems
were studied in MEEFT.
A remark is in order here on
how the short-range contributions are taken into account in MEEFT.
The basic premise of EFT is that physics pertaining to ranges shorter than the
experimentally relevant scale is to be lodged in local operators.
This means that, provided that a proper renormalization procedure is implemented to
the coefficients of the local operators ({\it i.e.,}  LECs),
the renormalization invariance ensures
that the net physical amplitudes be
independent of the details of short-range physics.
We implement the renormalization condition here
by adjusting the values of
LECs ($g_{4s}$ and $g_{4v}$ for $hen$ and $\dR$ for $hep$)
so as to reproduce a set of known experimental data
[ $\mu(\H3)$ and $\mu(\He3)$ for $hen$ and
the tritium-beta-decay rate for $hep$ ].
This matching procedure
should be done for each cutoff value and for each
potential adopted.
%
Differences in short-range contributions
calculated for each case
shift the values of LECs (which are not physical observables),
but the physical amplitudes should remain unaffected
if renormalization invariance is to hold.
The validity of this scheme can be checked
by monitoring the stability of the relevant physical observables
with respect to changes in the cutoff parameter $\Lambda$.
It turns out (see below) that, in the present $hen$ case,
the inclusion of the local-operator (or contact term) contributions
reduces the $\Lambda$-dependence
by a factor of $\sim$5,
demonstrating the
validity of the adopted renormalization procedure.
The residual $\Lambda$-dependence may be
ascribed to higher order contributions.

One might also worry about the current conservation of MEEFT
at short-range.
%
%
Although the M1 operators arise from
the transverse parts of the currents
(which by definition have vanishing divergence),
current conservation is still relevant in the present context.
The reason is that, 
since most of the Feynman diagrams 
generate both longitudinal and transverse parts simultaneously,
the current conservation breaking in the longitudinal part
of the calculated current signals possible mismatch in the M1 operator.
This can be part of the aforementioned mismatch problems in MEEFT.
The consequences of current conservation breaking
were studied by
Pastore et al. and Kolling et al.
\cite{Pastore:2008sk,Kolling:2009iq,Pastore:2009is}
and, according to these works,
current conservation violation has only minor effects.
We have also compared our current operators with those given in these references,
which use slightly different power counting schemes.
We have found that, after taking into account the renormalization for the LECs
and the fact that the contact terms are effective only at $S$-waves,
there is no difference in the two-pion and shorter-ranged contributions.
In the long-range region, however, there are additional Sachs terms in Ref.\cite{Girlanda:2009ct,Girlanda:2010vm}
whose coefficients are fixed by the nuclear Hamiltonian.
\footnote{After the submission of our paper, a full EFT calculation
of the HEN process became available\cite{Girlanda:2009ct,Girlanda:2010vm}.
The numerical results in this latest calculation are close
to those of our work here, despite differences in the formalisms.
}
Although the omission of those terms results in a violation 
of exact current conservation and leads to certain formal mismatches 
between the current structure and the potential, 
it is beyond the scope of the present work to solve these problems fully.
Here we take the viewpoint that the numerical consequences 
of these mismatches can be inferred from the cutoff-dependence 
of the calculated values of observables.  
It is reassuring that, in our case, this cut-off dependence turns 
out to be very weak (see Table \ref{tab:be_M_hen_cutoff}).

In the above we have focused on the short-range contributions.
It is however important to note
that, for $A$-body systems with $A\ge 3$,
even a so-called realistic nuclear interaction often fails
to reproduce accurately
the ERPs that govern the long-range
part of the transition matrix elements.
If such a mismatch in the long-range region occurs,
it cannot be cured by the renormalization of local operators,
a problem that can seriously affect the reliability
of a calculated transition amplitude.
As an exception to this general statement, however,
we should mention that,
if a clean correlation between the ERPs and the transition amplitude
under consideration can be established,
this correlation can be used to drastically reduce
the model dependence of the calculated transition amplitude~ \cite{slp2}.
This point will be explained in more detail later (Sec. III A).
Here we simply state that, by taking advantage of this feature,
we  obtain, as the best estimates for the threshold $hen$ cross section,
$\sigma= (49.4\pm 8.5)\ \mu b$ (for the AV18+UIX  potential),
and $(44.4\pm6.7)\ \mu b$ (for the I-N3LO+UIX$^*$ potential);
see eq.(\ref{sigma-hen}) for details.
Good agreement of these estimates with the experimental value of
the $hen$ cross section
gives strong support for the previous MEEFT calculation
for the $hep$ $S$-factor~\cite{hep}.

We wish to emphasize that
the present work is the first calculation of $hen$
that employs fully realistic nuclear wave functions
\footnote{After submission of our manuscript,
another EFT calculation of the $hen$ process
has appeared\cite{Girlanda:2009ct},
where both the potential and current operator are derived
in EFT.
};
these wave functions are numerically exact solutions
to the Faddeev-Yakubovsky equations in configuration space
for a specified realistic nuclear interaction.
It is to be noted that hitherto even the most advanced realistic
calculations~\cite{Carlson:1990nh,Schiavilla:1992sb,Song:2003ja}
 disregarded  the coupling of the $n$-$\He3$ to the $p$-$\H3$ state
in the asymptote of the initial  wave function.
This can have significant numerical consequences
in evaluating the 1B contributions;
see Sec.~\ref{comparison}, for details.
%

This paper is organized as follows.
In Sec. II we explain the formalisms used to
derive the M1 operators and to obtain the four-body nuclear wave functions.
Sec. III gives numerical results and analyses.
In the final section the implication of our work is discussed.

\section{Formalism}

\subsection{Electromagnetic M1 operators and the $hen$ cross section}

In this section we present M1 operators that arise from
the multipole expansion of the electromagnetic (EM) currents
obtained from HBChPT up to $\nlo3$ in chiral order counting.
HBChPT contains nucleons and pions as pertinent degrees of
freedom, with all the other massive fields integrated out.
In HBChPT, the EM currents
(and consequently the M1 operators) are expanded
systematically with increasing powers of $Q/\Lambda_\chi$, where
$Q$ stands for the typical momentum scale of the process and/or
the pion mass;
$\Lambda_\chi\sim 4\pi f_\pi \sim m \sim 1$~GeV
is the chiral scale, where $f_\pi\simeq 92.4$ MeV is the pion decay
constant, and $m$ is the nucleon mass. We remark
that, while the nucleon momentum $\vp_i$ is of the order of $Q$,
its energy ($\sim \vp_i^2/m$) is  of the order of
$Q^2/m$, and consequently the four-momentum of the emitted photon
$q^\mu=(\omega,{\bm q})$ should also be counted as ${\cal O}(Q^2/m)$.

We derive the MEC operators from the non-relativistic reduction
of irreducible contributions of Feynman diagrams in HBChPT.
Irreducible contributions coming from box diagrams are obtained
by removing pure nucleon-pole contributions.
As mentioned,
it is to be noted that
there exist other approaches to deriving MEC operators from HBChPT
\cite{Pastore:2008sk,Kolling:2009iq,Pastore:2009is}.
Although detailed comparison of our formalism with these approaches
should be informative, we relegate it to future studies.

The M1 operator $\vmu_{1M}(q)$ is defined as
\be
\vmu_{1M}(q)&\equiv& \left(\frac{iq}{\sqrt{6\pi}}\right)^{-1}{\hat T}^{Mag}_{1M}(q)
\ee
with
\bea
{\hat T}^{Mag}_{JM}(q)&\equiv&
\int\! d^3{\bm x}\, \left[j_J(qx){\bm Y}^M_{JJ1}(\hat{\bm x})\right]\cdot{\bm j}({\bm x}),
\eea
where $q\equiv|{\bm q}|=20.578$ MeV,
$j_J(qx)$ is the spherical Bessel function of order $J$,
${\bm Y}^{M}_{JJ1}(\hat{\bm x})$ is the vector spherical harmonics,
and ${\bm j}({\bm x})$ is the EM current operator.
We have chosen the normalization of $\vmu_{1M}(q)$
such that it becomes the usual magnetic dipole moment
in zero $q$ limit.

In terms of 
$\vmu_{1M}(q)$,
the $hen$ cross section at thermal energy is given by
\begin{equation}
\sigma=\alpha \pi \frac{c}{v_{rel}}
\left( \frac{q}{mc^{2}}\right)^{2}
\left( \frac{q}{\hbar c}\right)
\left\vert
\calM
\right\vert^{2}
\end{equation}%
with
\begin{equation}
\calM\equiv
\left\langle \Psi_{{}^{4}\rm{He}}^{J=0,M=0}
\right\vert \vmu_{11}(q)
\left\vert \Psi_{n^{3}\rm{He}}^{J=1,M=-1} \right\rangle,
\label{eq:calM}
\end{equation}%
where $\alpha $ is the fine structure constant, $m$
is the nucleon mass and
$v_{rel}=2200$ $m/s$ is the thermal neutron velocity.

The detailed full forms of the
M1 operators up to $\nlo3$
are given in our recent papers\cite{slp1,slp2},
which we briefly summarize here.
The M1 operators up to $\nlo3$ consist of one-body(1B)
and two-body(2B) contributions;
three-body operators enter only at \nlo4 or higher orders
in our counting scheme.

The 1B M1 operators read, in the center of mass frame,
\be
\vmu_{\rm 1B}(q) &=& \frac{1}{2 m} \sum_i  \Big\{
  \hat j_0(q r_i) \left[
 \vs_i \left( \mu_i - Q_i \frac{\vbp_i^2}{2 m^2}\right)
  -\frac{\mu_i-Q_i}{2 m^2}  \bar{\vp}_i\vs_i\cdot \bar{\vp}_i \right]
 \nonumber \\
 &&+\ \hat j_1(q r_i) \left[
 Q_i \vr_i\times \vbp_i \left(1 - \frac{\vbp_i^2}{2 m^2}\right)
 -\frac{w(2\mu_i-Q_i)}{4m} i\vr_i \times (\bar{\vp}_i\times\vs_i) \right]
 \nonumber\\
 &&+\ \frac{(q r_i)^2}{30} \hat j_2(q r_i)\, \mu_i\,
  (3 \hat \vr_i \,\hat \vr_i\cdot \vs_i - \vs_i)
 + \cdots \Big\},
\label{vmu1B}\ee
where
$\hat j_n(x)\equiv \frac{(2n+1)!!}{x^n} j_n(x)$,
$Q_i$ and $\mu_i$ are the charge and magnetic moments of the $i$-th nucleon,
respectively,
and $\vbp_i\equiv \frac{1}{2}(i\overleftarrow{\nabla}_i-i\overrightarrow{\nabla}_i)$
is the mean momentum operator of the $i$-th nucleon.

The two-body M1 operators
\footnote{
In this work we neglect
the so-called fixed-current contribution,
which is proved to be numerically negligible~\cite{slp1}.
}
up to \nlo3 can be divided into four types:
the {\em soft}-one-pion-exchange ($1\pi$) term,
the vertex correction to the one-pion exchange $(1\pi C)$ term,
the two-pion-exchange $(2\pi)$ term, and
the contact-term contribution (CT),
\be \vmu_{\rm 2B}(q) = \sum_{i<j} \left[\vmu_{ij}^{1\pi} +
\left(\vmu_{ij}^{1\pi C}+\vmu_{ij}^{2\pi} + \vmu_{ij}^{\rm CT}\right)\right] = \nlo{} + \nlo3.
\ee
The {\em soft}-one-pion-exchange ($1\pi$) term is NLO and can be
written in terms of
$\vr=\vr_j-\vr_k$, $r=|\vr|$, $\hatr={\bm r}/|\vr|$,
$\vR=(\vr_j+\vr_k)/2$, $R=|\vR|$,
$S_{jk}=3\vs_j\cdot\hatr\vs_k\cdot\hatr-\vs_j\cdot\vs_k$,
$g_A\simeq 1.2695$,
\be
\vmu_{jk}^{1\pi} &=& \frac{g_A^2}{8f_\pi^2}
  \left[{\hat T}_{S}^{(\times)}
  \left(\frac{2}{3}y_{1\Lambda}^\pi(r)-y_{0\Lambda}^\pi(r)\right)
        -{\hat T}_{T}^{(\times)}y_{1\Lambda}^\pi(r)\right]
   \hat j_0 (q R)
\no &-&\frac{g_A^2 m_\pi^2}{24 f_\pi^2}\tau_{\times}^z
     \vR \times \vr
          \left[\vs_1\cdot\vs_2 {\bar y}_{0\Lambda}^\pi(r)
           +S_{jk} y_{2\Lambda}^\pi(r)\right]
          \hat j_1 (q R)
      + \cdots,
\label{1pi}
\ee
where,
${\hat T}_{S}^{(\odot)}\equiv \tau_\odot^z
\vs_\odot$ and ${\hat T}_{T}^{(\odot)}\equiv \tau_\odot^z
  \left[\hatr\,\hatr\cdot\vs_\odot-\frac13\vs_\odot\right]$,
$\tau_{\odot}=\tau_1\odot\tau_2$, $\vs_{\odot}=\vs_1\odot\vs_2$,
with $\odot= \pm,\, \times$. The cutoff dependence of 2B
operators is taken into account
by introducing a Gaussian regulator with a cutoff $\Lambda$
when performing Fourier transformation
of the 2B operators into coordinate space;
this procedure gives the
regularized delta and Yukawa functions,
\bea
\delta_\Lambda(r)&\equiv&\int\!\! \frac{d^3 \vk}{(2\pi)^3}
\,e^{-\vk^2/\Lambda^2}\,
e^{i\vk\cdot\vr}, \no
y_{0\Lambda}^\pi(r)&\equiv&\int \frac{d^3 \vk}{(2\pi)^3}
\,e^{-\vk^2/\Lambda^2}\,
e^{i\vk\cdot\vr}
\frac{1}{\vk^2+m_\pi^2},\no
y_{1\Lambda}^\pi&\equiv& -r\frac{d}{dr}y_{0\Lambda}^\pi,\quad
y_{2\Lambda}^\pi\equiv\frac{r}{m_\pi^2}
\frac{d}{dr}\frac{1}{r}\frac{d}{dr} y_{0\Lambda},
\eea
where, $m_\pi$ is the pion mass.

The one-loop vertex correction to one-pion exchange
($1\pi C$ term) reads
\be
\vmu_{jk}^{1\pi C} &=& -\frac{g_A^2}{8 f_\pi^2}
  ({\bar c}_\omega+{\bar c}_\Delta)
  \left[({\hat T}_{S}^{(+)}+ {\hat T}_{S}^{(-)})
  \frac{{\bar y}^\pi_{0\Lambda}}{3}
  +({\hat T}_{T}^{(+)} + {\hat T}_{T}^{(-)})\,y^\pi_{2\Lambda}\right]
  \hat{j}_1(qR)
\no
  &&
+\frac{g_A^2 }{8 f_\pi^2} {\bar c}_{\Delta}
 \left[\frac13\TtimesS{\bar y}^\pi_{0\Lambda}
   -\frac{1}{2}\TtimesT y^\pi_{2\Lambda}\right]
  \hat j_1(qR)
\no &&-\frac{1}{16 f_\pi^2}{\bar N}_{WZ}\tau_j\cdot\tau_k
  \left[\vs_{+}{\bar y}^\pi_{0\Lambda}
+(3\hatr\hatr\cdot\vs_{+}-\vs_{+}) y^\pi_{2\Lambda}\right]
 \hat j_1(qR),
\label{vmu1piC}\ee
where the values of the LECs,
$(\bar c_\omega,\ \bar c_\Delta,\
{\bar N}_{WZ})\simeq (0.1021,\ 0.1667,\ 0.02395)$, are determined
from the resonance saturation model\cite{npdg,M1S}.

The two-pion exchange ($2\pi$) term reads
\be \vmu_{jk}^{2\pi} &=& \frac{1}{128\pi^2
f_\pi^4}\left[\left(\TplusS-\TminusS\right)L_S(r)
              +\left(\TplusT-\TminusT\right)L_T(r)\right]
   \hat j_1(qR)
\no &-& \frac{1}{256\pi^2 f_\pi^4}(\tau_j\times\tau_k)^z \vR \times \hatr
\frac{d}{dr} L_0(r) \hat j_0(qR)
\, , \label{2pi}\ee
where
\bea
L_S(r) &=& -\frac{g_A^2}{3} r \frac{d}{dr} K_0  +
\frac{g_A^4}{3} \big[ 4 K_1 -2 K_0+r\frac{d}{d r} (K_0 + 2 K_1)
\big], \no
L_T(r)&=&  \frac{g_A^2}{2}  r\frac{d}{d r} K_0 +
\frac{g_A^4}{2} \big[ 4 K_{\rm T} - r \frac{d}{d r} (K_0 + 2 K_1)
\big], \no
L_0(r) &=&2 K_2 + g_A^2 (8K_2+2K_1+2K_0) \no && \ \ \
        -g_A^4 (16 K_2+5 K_1+5 K_0)+ g_A^4 \frac{d}{dr} (r K_1).
\eea
The loop functions $K's$ are defined in Ref. \cite{hep, npdg} .

Note that the contact-term  $\vmu_{ij}^{\rm CT}$ contains two
low-energy constants (LECs), $g_{4s}$ and $g_{4v}$:
\be
\vmu^{\rm CT}_{ij} &=& \frac{1}{2m}
         \left[g_{4s}(\vs_i+\vs_j)
         +g_{4v} (\vec{\tau}_i\times \vec{\tau}_j)^z (\vs_i \times \vs_j)\right]
         \delta^{(3)}_\Lambda(\vr_{ij}).
\label{CT}
\ee
Both $g_{4s}$ and $g_{4v}$ have the dimension of $\fm^3$.
Since the values of these LECs are not determined by symmetry arguments,
they need to be fixed either by solving QCD at low-energy
or by fitting to a set of experimental observables that are sensitive to them.
Since the former is not feasible at present,
we resort to the latter.
Specifically,
we fix $g_{4s}$ and $g_{4v}$ so as to reproduce
the experimental values of $\mu(\H3)$ and $\mu(\He3)$,
for each nuclear interaction model adopted and for each cutoff value.

\subsection{Faddeev-Yakubovsky equations \label{sec_FY_eq}}

\begin{figure}[h!]
\begin{center}
\mbox{\epsfxsize=7.6cm\epsffile{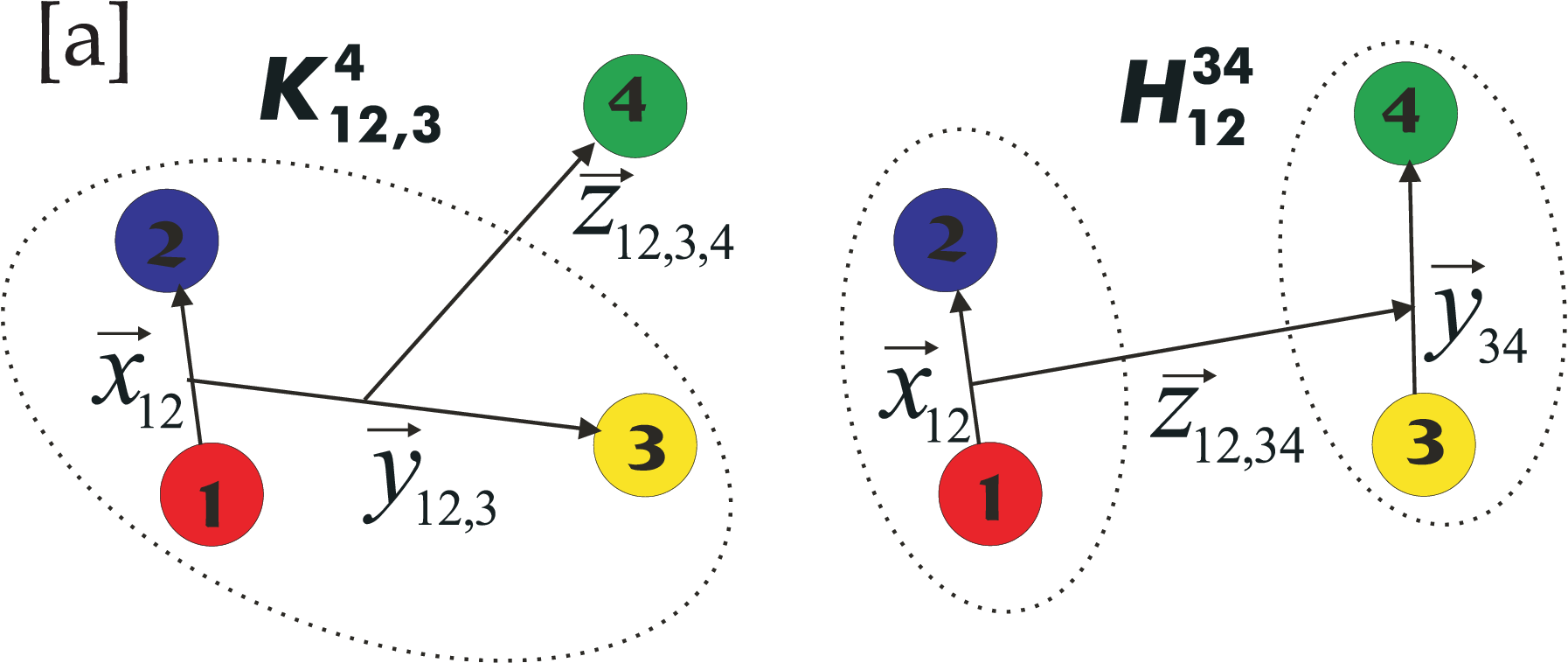}}\hspace{0.2cm} %
\mbox{\epsfxsize=8.4cm\epsffile{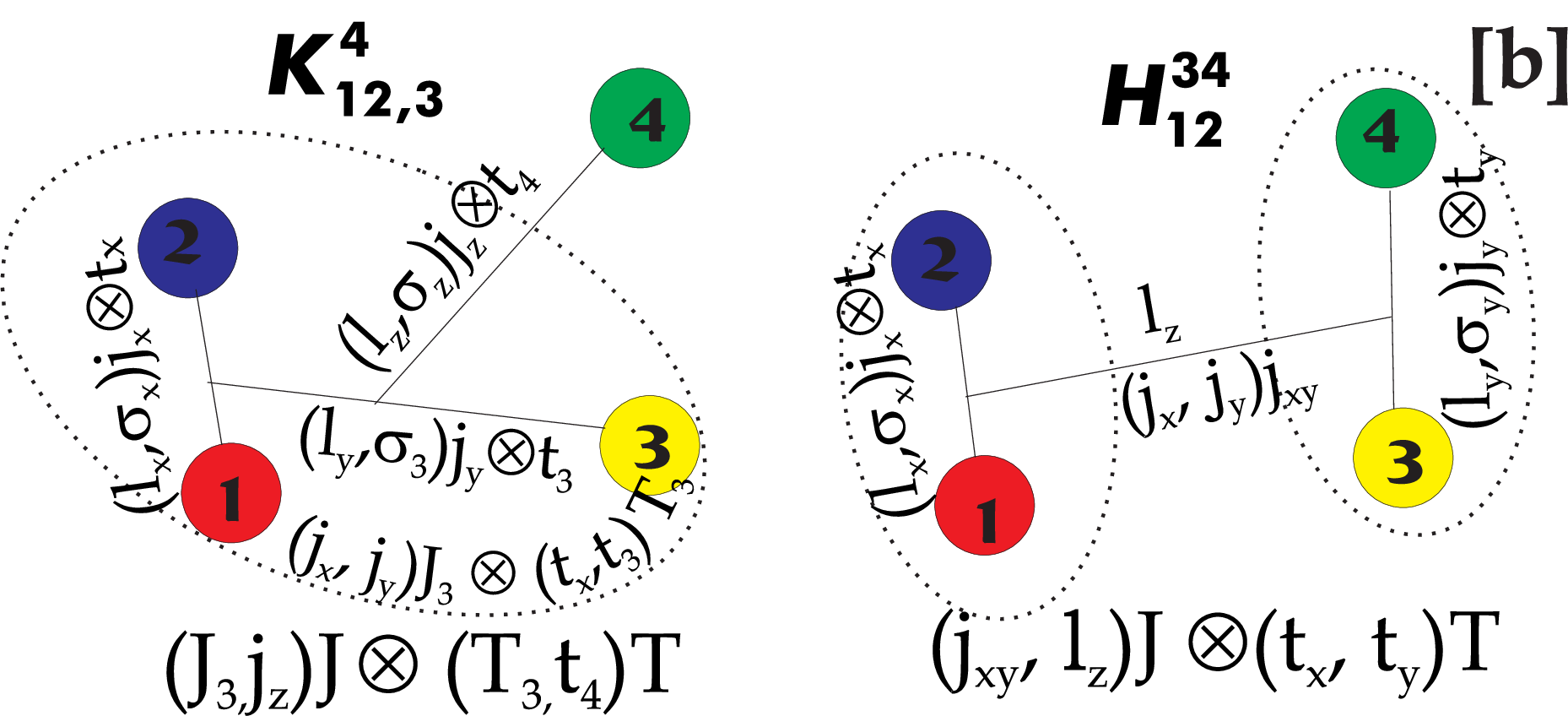}}
\end{center}
\caption{
(Color online)
The FY components $K_{12,3}^4$ and $H_{12}^{34}$ for a given particle ordering. As $z\rightarrow \infty$,
the $K$ components describe 3+1 particle channels,
while the $H$ components contain
asymptotic states of 2+2 channels, see figure~[a].
Figure~[b] shows the \textit{j-j}
coupling scheme used in expanding $K$ and $H$ into partial wave
bases.}
\label{Fig_4b_config}
\end{figure}

The relevant four-nucleon wave functions are obtained
by solving the Faddeev-Yakubovsky (FY)
equations in configuration space~\cite{Fadd_art,Yakub}.
The FY formalism offers a mathematically rigorous description
for both continuum and bound states.
In this formalism wave functions are naturally decomposed into so-called
FY amplitudes (FYAs).
For $A=4$ systems, there appear two types of FYAs, which we refer to as
components $K_{ij,k}^l$ and $H_{ij}^{kl}$
where $i,j,k,l$ are particle indices.
The asymptotes of the components $K_{ij,k}^l$ and $H_{ij}^{kl}$
 incorporate 3+1 and 2+2 particle channels, respectively (see Fig.\ref{Fig_4b_config}).
The FYAs are evaluated in the isospin formalism,
{\it i.e.},  protons and neutrons
are regarded as degenerate states with the same mass,
which is
fixed to $\hbar ^{2}/m=41.47$ MeV$\cdot$fm$^2$.
Three-body forces typically arise from integrating out
the higher-energy degrees of freedom,
and therefore they can be decomposed as
 $V_{123}=V_{12}^3+V_{23}^1+V_{31}^2$,
where $k$ in $V_{ij}^k$ is the particle 
in a high-energy intermediate state.\cite{POT_UIX}.
In the presence of a three-body force,
the FY equations for
$K\equiv K_{12,3}^4$ and $H\equiv H_{12}^{34}$
read~\cite{These_Rimas_03,Lazauskas:2008kz}
\begin{eqnarray}
\left( E-H_{0}-V_{12}-\sum_{i<j}V_{ij}^{C}\right) K
&=&V_{12}(P^{+}+P^{-})\left[ (1+Q)K+H\right]
+\frac{1}{2} \left(V_{23}^1+V_{31}^2\right) \Psi ,
\notag \\
\left( E-H_{0}-V_{12}-\sum_{i<j}V_{ij}^{C}\right) H
&=&V_{12}\tilde{P}\left[ (1+Q)K+H\right] ,
\label{FY1}
\end{eqnarray}%
where $V_{ij}$ and $V_{ij}^{C}$ are,
respectively,  the short-ranged part
and the Coulomb-dominated long-range part of the interaction
between the $i$-th and $j$-th nucleons.
$P^{+}=(P^{-})^{-1}\equiv P_{23}P_{12}$,
$Q\equiv - P_{34}$ and $\tilde{P} \equiv P_{13}P_{24}=P_{24}P_{13}$,
where $P_{ij}$ is the particle permutation operator.
In terms of the FYAs, the total wave function of an $A=4$ system
is given by
\begin{equation}  \label{FY_wave_func}
\Psi =\left[ 1+(1+P^{+}+P^{-})Q\right]
(1+P^{+}+P^{-})K +(1+P^{+}+P^{-})(1+\tilde{P})H.
\end{equation}
We expand $K$ and $H$ in terms of the
tripolar harmonics
$Y_{i}^{\alpha }(\hat{x}_{i},\hat{y}_{i},\hat{z}_{i})$,
which comprise
the spins and isospins of the nucleons as well as the angular variables,
\begin{equation}
\Phi_{i}(\vec{x}_{i},\vec{y}_{i},\vec{z}_{i})
=\sum_{\alpha }
\frac{\mathcal{F}_{i}^{\alpha }(x_{i},y_{i},z_{i})}{x_{i}y_{i}z_{i}}
Y_{i}^{\alpha }(\hat{x}_{i},\hat{y}_{i},\hat{z}_{i}),
\label{PWB_decomp}
\end{equation}
where $\Phi$ stands for either $K$ or $H$,
and the subscript $i$ denotes the particle-grouping class (among the four nucleons).
We note that
the total angular momentum and its projection, parity and the
third component of the isospin ($\mathcal{T}_z = 0$)
are good quantum numbers,
and the subscript $\alpha$ denotes collectively
eleven other non-fixed quantum numbers.
We use the $j$-$j$ scheme for the coupling of angular momenta,
as illustrated in Fig.~\ref{Fig_4b_config}$b$.
The Jacobi coordinates used here
are depicted in Fig.~\ref{Fig_4b_config}.
This choice of coordinates
allows us to separate the center-of-mass motion and
guarantees that the
kinetic energy operator is independent of
the angular variables.

The expansion of Eq.~(\ref{FY1}) in terms of the
natural configuration space basis leads to
coupled integro-differential equations for the
radial parts of FYAs ($\mathcal{F}_{i}^{\alpha }(x_{i},y_{i},z_{i})$).
Note that, contrary to the ordinary 3$N$ problems,
the number of radial parts of FYAs is infinite
even when the pair interaction is restricted to a
finite number of partial waves.
This situation arises from the existence of the
additional degree of freedom $l_{z}$ in the expansion of the $K$-type
components.
In numerical calculations, therefore,
we need to introduce an additional truncation
by identifying relevant amplitudes and discarding the remainder
(see below).

\subsection{Boundary conditions}

Eqs. (\ref{FY1})
needs to  be supplemented with appropriate boundary conditions,
which can be written in the Dirichlet form.
For both bound and scattering states,
the radial FYAs satisfy the regularity conditions:
\begin{equation}
\mathcal{F}_{i}^{\alpha}(0,y_{i},z_{i})=
\mathcal{F}_{i}^{\alpha}(x_{i},0,z_{i})=
\mathcal{F}_{i}^{\alpha}(x_{i},y_{i},0)=0.
\label{BC_xyz_0}
\end{equation}
For bound state problems, since the wave functions are compact, the
regularity conditions can be implemented by requiring
$\mathcal{F}_{i}^{\alpha }$ to vanish at the borders of the hypercube,
$\left[ 0,X_{\max }\right] \times \left[ 0,Y_{\max }\right] \times \left[ 0,Z_{\max }\right]$,\footnote{
%
$(X_{\max},\, Y_{\max},\,Z_{\max})$ are chosen to be
$(L_y,\, \sqrt{3/4} L_y,\, \sqrt{2/3} L_z)$ for the component K
and $(L_y,\, L_y,\, \sqrt{1/2} L_z)$ for the component H,
where $L_y= 25\ \fm$ and $L_z= (27\sim30)\ \fm$.
We have verified that
a hypercube size larger than these values
does not cause any noticeable changes.
}
\begin{equation}
\mathcal{F}_{i}^{\alpha}(X_{\max},y_{i},z_{i})
=\mathcal{F}_{i}^{\alpha}(x_{i},Y_{\max},z_{i})
=\mathcal{F}_{i}^{\alpha}(x_{i},y_{i},Z_{\max})=0.
\label{BC_BS}
\end{equation}
The hypercube is chosen large enough to accommodate
the wave functions.

On the other hand, a scattering state near threshold
contains two coupled channels, $n$-$\He3$ and $p$-$\H3$,
both of which are of type $K$.
In this case
we impose the following matching condition at $z_{i}=Z_{\max}$:
\begin{eqnarray}
\mathcal{K}_{i}^{\alpha }(x_{i},y_{i},Z_{\max })
&=&\frac{1}{\sqrt{4}} \sum_{j_{z}^{\prime} l_{z}^{\prime}T_{3}^{\prime z}}
\left. \left\{
f_{i}^{\alpha _{a}}(\vx_{i},\vy_{i})\right\}_{J_{3}\equiv\frac{1}{2},T_{3}T_{3}^{z}}
\otimes \left\{ Y_{l_{z}^{\prime }}(\widehat{z}_{i})\otimes s_{i}
\right\}_{j_{z}^{\prime }}\right\rangle_{JM}
\notag \\
&\times &\left( \vphantom{\sqrt{\frac{p'}{p}}}
\frac{i}{2}\left[ \delta_{l_{z},0}
h_{l_{z}}^{-}(p_{n}Z_{\max})-S_{j_{z}^{\prime }l_{z}^{\prime}
T_{3}^{\prime z},j_{z}l_{z}T_{3}^{z}}h_{l_{z}^{\prime }}^{+}(p_{n}Z_{\max })\right]
C_{T_{3}T_{3}^{\prime z},\frac{1}{2}-\frac{1}{2}}^{T0}
\right.
\notag \\
&&\left.-\
\frac{i}{2}\sqrt{\frac{p_{p}^{\prime }}{p_{n}}}
S_{j_{z}^{\prime}l_{z}^{\prime}T_{3}^{\prime z},j_{z}l_{z}T_{3}^{z}}
e^{2i\sigma_{l_{z}^{\prime }}}
u_{l_{z}^{\prime }}^{+}(\eta ,p_{p}^{\prime }Z_{\max})
\,C_{T_{3}T_{3}^{\prime z},\frac{1}{2}\frac{1}{2}}^{T0}
\right).
\label{eq_as_beh}\end{eqnarray}%
Here
$p_{n}=\frac{3}{4\hbar} m v_{rel}$ is the
neutron  momentum in the $n$-$\He3$ channel,
while $p_{p}=\left[p_{n}^{2}+\frac{3m}{2\hbar ^{2}}
\left( B_{{}^3\rm{H}}-B_{{}^{3}\rm{He}}\right) \right]^{1/2}$
is the proton momentum
in the $p$-$\H3$ channel;
$h_{l_{z}}^{\pm }$ are
the spherical Hankel functions,
and
$u_{l_{z}^{\prime }}^{+}$ is the outgoing Coulomb function
for the $p$-$\H3$ channel
with $\eta =\frac{4}{3}\alpha mc/(\hbar p_{p})$.
The functions
$\left\{ f_{i}^{\alpha _{a}}(\vx_{i},\vy_{i})
\right\}_{J_{3}\equiv \frac{1}{2},T_{3}T_{3}^{z}}$
with $T_{3}^{z}=\pm \frac{1}{2}$ are the normalized
Faddeev amplitudes for $\He3$ and $\H3$,
which we obtain by solving the corresponding $3N$ bound state problems.
%
%

We neglect the $\mathcal{T}\neq 0$ components in our $\He4$ bound-state calculation as they represent less than 0.01\% \cite{Nogga_alp} of the total wave function.
However, for the rigorous solution of the scattering problem,
we cannot use the $\mathcal{T}=0$ approximation;
we do need to consider admixtures of
$\mathcal{T}= 1$ and $\mathcal{T}=2$ states.
These admixtures are needed in order 
to correctly separate the asymptotes of the
$n$-$\He3$ and $p$-$\H3$ channels,
which have different thresholds due to the difference 
in the $\He3$ and $\H3$ binding energies.

\section{Results}

In this work we have performed rigorous FY calculations
for five sets of nuclear potentials:
AV18, I-N3LO, INOY, AV18+UIX and I-N3LO+UIX*.
Here AV18 stands for
the Argonne v18 potential~\cite{POT_AV18},
I-N3LO for the chiral \nlo3 potential of the Idaho group~\cite{POT_IN3LO},
and INOY for the non-local configuration space potential
that has recently been derived by Doleschall~\cite{POT_INOY};
UIX is the tri-nucleon interaction derived by the Urbana group~\cite{POT_UIX}.
For the case of I-N3LO+UIX*,
we have attached an asterisk to indicate that
the $A_{2\pi}$ parameter in UIX has been slightly modified
(from $-0.0293$ to $-0.03827$)
so as to reproduce the triton binding energy precisely, see Ref.~\cite{slp2}
for details.


For each of these five nuclear interactions,
and for each of the three choices of the cutoff parameter,
$\Lambda$ = 500, 700 and 900 MeV,
we determine the LECs, $g_{4s}$ and $g_{4v}$,
in such a manner that the experimental values of
the triton and $^3$He magnetic moments
are reproduced.
We then proceed to calculate the $hen$ cross section,  $\sigma$.
The results are given in Table~\ref{tab:be_sigma_hen}.
The table also shows the calculated values of
the binding energies of $\H3$, $\He3$ and $\He4$,
the point-proton rms radius $r_{\rm{He4}}$ of $\He4$,\footnote{
\protect
The point-proton rms radius $r_{\rm{He4}}$ is defined as
$(r_{\rm{He4}})^2 \equiv r_c^2(\He4)  - r_p^2 - r_n^2$,
where $r_c(\He4)$ is the proton charge rms radius of $\He4$,
$r_p$ and  $r_n$ are the rms charge radius of the proton and neutron, respectively.
See, for example, \cite{rms} for detailed explanation.
With the $\He4$ proton charge radius
$r_c(\He4)= 1.681(4)\ \fm$ obtained in a recent analysis~\cite{ingo}
and with the 2008 PDG values for the proton and neutron rms radii,
$r_p= 0.875(7)\ \fm$
and $r_n^2= -0.1161(22)\ \fm^2$, we arrive at
$r_{\rm{He4}} = 1.475(6)\ \fm$, which is about 1.4\ \% larger than
the estimate given in ~\cite{rms}.
}
the $D$-state probability of $\He4$,
and the spin-triplet $n$-$\He3$ scattering length, $a_{n\rm{He3}}$.
\begin{table}[tbp]
\caption{\label{tab:be_sigma_hen}
\protect The $hen$ cross section, $\sigma$,
calculated for the five realistic nuclear interactions
mentioned in the text.
The uncertainties attached to $\sigma$ represent the variation
of $\sigma$ as the cutoff parameter $\Lambda$ is varied
in the range $\Lambda = (500 \sim 900)$ MeV.
Also shown are the calculated values of
the binding energies (BE; in units of MeV)
for $\H3$, $\He3$ and $\He4$,
the point-proton rms radius~\cite{ingo} $r_{\rm{He4}}$ (in fm),
the $D$-state probability $P_D(\He4)$ (in per cent)
of the $\alpha$-particle, and 
the spin-triplet $n$-$\He3$ scattering length, $a_{n\rm{He3}}$
(in units of fm).
}
\begin{ruledtabular}
\begin{tabular}{lccccccc}
            & BE($\H3$) & BE($\He3$) & BE($\He4$)& $r_{\rm{He4}}$  & $P_D(\He4)$ &
$a_{n\rm{He3}}$ & $\sigma$ [$\mu b$] \\\hline
AV18 & 7.623 & 6.925 & 24.23& 1.516 & 13.8 & $3.43-0.0082i$ & $80.0 \pm 12.2$ \\
I-N3LO & 7.852 & 7.159 & 25.36&1.52 & 9.30 & $3.56-0.0070i$ & $57.3 \pm 7.9$ \\
INOY & 8.483 & 7.720 & 29.08& 1.377 & 5.95 & $3.26-0.0058i$ & $34.4\pm 4.5$ \\
AV18+UIX & 8.483 & 7.753 & 28.47&1.431 & 16.0 & $3.23-0.0054i$ & $49.4\pm 8.5$ \\
I-N3LO+UIX* & 8.482 & 7.737 & 28.12& 1.475 & 10.9 & $3.44-0.0055i$ & $44.4\pm6.7$ \\\hline
Exp.: & 8.482 & 7.718 & 28.30 & 1.475(6) & & $3.278(53)-0.001(2)i$& $55\pm3$, $54\pm 6$
\end{tabular}
\end{ruledtabular}
\end{table}

The bound state properties calculated in this work
agree well with those obtained in
other calculations~\cite{Arnas_be,Kievsky:2008es,Nogga:2000uu}.
Theoretical calculations for $a_{n\rm{He3}}$
are much less established,
but we have checked
that our results agree within 2$\%$ with the
momentum-space FY calculation of the Lisboa group~\cite{Arnas_pr},
as well as with the RGM calculation
carried out  by Hofmann~\cite{Hofmann:2002xr}
for the AV18 and AV18+UIX potentials.

The table also indicates that the three-nucleon interactions (TNIs)
play an important role in
bringing the binding energies and $a_{n\rm{He3}}$
close to their respective experimental values.
We remark that there is some uncertainty in the experimental value of
$a_{n\rm{He3}}$.
The value listed in the table is
due to an $R$-matrix analysis~\cite{Huber:2008kc}
of the $n$-$\He3$ scattering data
measured before the year 2002.
Recently new measurements of the coherent scattering length
have been performed at
NIST~\cite{Huffman:2004wv} and ILL~\cite{Ketter:2006},
but the results of the two groups do not agree with each other,
and both of them are
in disagreement with the old ILL measurement.

The $D$-state probability of the $\alpha$-particles, $P_D(\He4)$,
which is closely related to the tensor forces,
shows strong model dependence.
However this quantity is not an observable,
and it turns out to be difficult to constrain this quantity
by studying other processes
that are sensitive to $P_D(\He4)$~\cite{Arriaga:1991zz}.

\begin{table}[tbp]
\caption{\label{tab:amp_conv}
\protect The real parts of contributions to the $hen$ amplitude
from the indicated types of the transition operators
calculated for three different values of $\mbox{max}(j_i)$,
the maximal value of partial angular momenta $j_i$
allowed in the expansion of FYAs.
We show here the results obtained
with the INOY wave functions and with $\Lambda$=700 MeV.
The one-body leading order (1B:LO)
contribution represents the impulse approximation terms,
while 1B:RC corresponds to the relativistic corrections to
the one-body current.
The finite-range two-body current is decomposed
into NLO one-pion exchange (${1\pi}$),
$\nlo3$ pion-loop corrections (${1\pi}C$)
and $\nlo3$ two-pion exchange (${2\pi}$) terms.
The contact-type two-body current is decomposed
into the $g_{4s}$ and $g_{4v}$ terms.
All the matrix elements are given in units of $\fm^{3/2}$.
For the values of $g_{4s}$ and $g_{4v}$
relevant to the present case, see Table III.
}
\begin{ruledtabular}
\begin{tabular}{lrrr}
 & max$(j_i)\leq$3 & max$(j_i)\leq$4 & max$(j_i)\leq$5 \\ \hline
\hline
1B: LO  &  $0.0455$ & $0.0496$ & $0.0511$ \\ 
1B: RC  &  $0.0537$ & $0.0535$ & $0.0534$ \\
1B-total&  $0.0992$ & $0.1031$ & $0.1045$ \\ \hline
2B: ${1\pi}$ (NLO)	    &$-0.0771$ & $-0.0781$ & $-0.0786$ \\	
2B: ${1\pi}C$ (\nlo3)	&$-0.0855$ & $-0.0861$ & $-0.0866$  \\	
2B: ${2\pi}$	(\nlo3)	&$-0.0380$ & $-0.0383$ & $-0.0384$\\ \hline
finite (total w/o CT)   &$-0.2006$ & $-0.2025$ & $-0.2035$\\ \hline
2B:$g_{4s}$             &$0.0471 g_{4s}$ &  $0.0472 g_{4s}$ &  $0.0473 g_{4s}$\\
2B:$g_{4v}$             &$-0.0718 g_{4v}$ & $-0.0722 g_{4v}$ & $-0.0725 g_{4v}$\\
2B: CT (\nlo3)   &$-0.2473$ &$-0.2495$ &$-0.2507$ \\ \hline
Total		&$-0.1482$& $-0.1464$ & $-0.1462$ \\
\end{tabular}
\end{ruledtabular}
\end{table}

As mentioned, to solve the FY equation numerically,
we need to introduce truncations
in the angular momentum expansion of the FYAs.
We implement these truncations
by assuming that the partial FYAs with $j_x$, $j_y$, $j_z$ (and $l_z$ for the type H)
larger than a specified value, $\mbox{max}(j_i)$,  can be ignored.
To illustrate the convergence property of the $hen$ amplitude
as a function of $\mbox{max}(j_i)$,
we show in Table \ref{tab:amp_conv}
the results obtained for $\mbox{max}(j_i)$= 3, 4 and 5.
The table gives the real parts of individual contributions from the indicated
types of the transition operators,
calculated with the INOY interaction for $\Lambda=700$ MeV.
%
One can see that, with $\mbox{max}(j_i)$ =5
(used in the present work),
the numerical accuracy of a few percent or better is achieved.
The leading one-body (1B:LO)\footnote{
Among the one-body M1 operator given in eq.(\ref{vmu1B}),
the 1B:LO corresponds to
$$
\vmu_{\rm 1B:LO}(q) = \frac{1}{2 m} \sum_i  \left[
  \mu_i  \hat j_0(q r_i) \vs_i
 + Q_i  \hat j_1(q r_i) \vr_i\times \vbp_i\right].$$
}
contribution shows the slowest convergence,
which can be understood by recalling that the 1B impulse contribution
undergoes a huge cancelation due to the
orthogonality
between the incoming and outgoing nuclear wave functions.
We have also checked that the convergence pattern is similar for the results with
other nuclear potentials as well as
for the imaginary parts of the transition amplitudes.

It turns out that
the 1B:RC is dominated by
the last term of the 2nd line of eq.(\ref{vmu1B}),
which accounts more than 95 \% of 1B:RC.
Note that this term is proportional to $\omega$ (the energy of the emitted photon),
and hence does not contribute to the magnetic-moment operators.

As can be seen from Table \ref{tab:amp_conv},
the calculated value of the $hen$ cross section, $\sigma$,
for AV18 (INOY) is too large (small) compared with
the experimental value, $\sigma_{exp}$,
while the results for the remaining three nuclear interactions
exhibit only mild variations around $\sigma_{exp}$.
A detailed discussion of this model dependence will be given
in the next subsection, but this is a good place to discuss
certain features specific to AV18 and INOY.
First, the conspicuous deviation of $\sigma$(AV18)
from $\sigma_{exp}$ is not surprising,
since AV18 without additional three-body nuclear interactions
fails to reproduce
the binding energies of the relevant nuclei.
We also remark that,
although the INOY potential is capable of reproducing
the binding energies and rms radii of the
$A=3$ system quite accurately~\cite{Lazauskas:2004hq},
it is known to give too large binding energies
and too small rms radii
for $A\ge 4$ nuclei~\cite{Forssen:2009vu} ;
it also gives overbound and too dense
nuclear matter~\cite{Baldo:2005jp}.
Such a feature leads to the reduction of the overlap 
between the $n$-$\He3$ and $\He4$ wave functions
and hence to underestimation of $\sigma$.
It is also to be noted that the results for INOY
deviate from the Tjon-line
(a line that correlates the $A=3$ and $A=4$ binding energies)
rather severely,
indicating that caution should be exercised
in using INOY for the $A\ge4$ systems.

For further discussion,
we list in Table~\ref{tab:be_M_hen700}
the values of the $hen$ matrix element $\calM$ [eq.(\ref{eq:calM})],
and the LECs,  $g_{4s}$ and $g_{4v}$,
evaluated at $\Lambda=700$ MeV,
for each of the five
nuclear interactions under consideration.
The real part of $\calM$ is written as the sum of
1B and 2B contributions,
with the dependence on $g_{4s}$ and $g_{4v}$ also shown.
\begin{table}[tbp]
\caption{\protect
Real and imaginary parts
of the $hen$ matrix element  $\calM$
(in units of $\fm^{3/2}$) calculated for $\Lambda=700$ MeV.
Also listed are the values of $g_{4s}$ and $g_{4v}$ (in units of $\fm^3$)
determined so as to reproduce the magnetic
moments of the $\H3$ and $\He3$ nuclei.
$\Re \calM$ is written in the format of:
(1B) + (2B w/o CT) + ($g_{4s}$-term) + ($g_{4v}$-term) = (total).
}
\label{tab:be_M_hen700}%
\begin{ruledtabular}
\begin{tabular}{crrcc}
   & $g_{4s}$ & $g_{4v}$ & $\Re \calM$ & $\Im \calM$ \\ \hline 
AV18 & 0.3958 & 0.1947 & $0.1531-0.3777 +0.0237\, g_{4s}-0.0403\, g_{4v}=-0.2231$ & 0.0249 \\
I-N3LO & 0.3919 & 2.7479 & $0.1304 -0.2248 +0.0198\, g_{4s}-0.0371\, g_{4v}=-0.1885$ & 0.0203 \\
INOY & 0.2313 & 0.8021 & $0.1045-0.2035 +0.0473\, g_{4s}-0.0725\, g_{4v}=-0.1462$ & 0.0154 \\
AV18+UIX & 0.5810 &$-0.4615$ & $0.1518-0.3567 +0.0205\, g_{4s}-0.0377\, g_{4v}=-0.1756$ & 0.0179\\
I-N3LO+UIX* & 0.5402 & 2.3249 & $0.1305-0.2253 +0.0175\, g_{4s}-0.0347\, g_{4v}=-0.1661$ & 0.0183 \\
\end{tabular}
\end{ruledtabular}
\end{table}
We see from the table that 
the 2B contributions are about two times as large as
the 1B contributions and that the 2B and 1B terms have opposite signs.
These features are consistent with the observation made
in~\cite{Carlson:1990nh,Song:2003ja}.
Secondly, there are substantial model dependence even in the 1B sector,
which might be traced to the fact that
not all the adopted nuclear potentials accurately reproduce the ERPs
that govern the long-$r$ contributions of 1B.
Finally, the inclusion of TNI(UIX) plays quite an important role
in reducing the model dependence.

\subsection{Model-dependence}

As mentioned, the calculated values of the $hen$ cross section
$\sigma$, shown in Table~\ref{tab:be_sigma_hen}
exhibit significant dependence on the nuclear interactions used.
In examining this model dependence,
it is informative to recall the results of
our previous MEEFT study~\cite{slp2} on  the M1 properties
of the $A\le 3$ nuclei.
It was found in~\cite{slp2} that the M1 matrix elements (MEs) of the
$A$=$3$ systems and the triton binding energy $B_3$
calculated for various realistic nuclear interactions
exhibit strong correlations and they lie
on a well-defined curve in the MEs-$B_3$ plot.
Meanwhile, since $B_3$ governs the long-distance contributions to the MEs,
the model-dependence
(viz., variations in the MEs corresponding to
the different nuclear potentials
that give different values for $B_3$)
cannot be cured by renormalizing the local (or short-ranged) operators.
As discussed in \cite{slp2}, however,
the use of the empirical correlation curve
between the MEs and $B_3$ allows us to drastically reduce
scatter in the calculated values of the MEs.
This is achieved by introducing a constraint that
only those values of the MEs be accepted which,
along the correlation curve,  have values of $B_3$
consistent with its experimental value.
This constraint was found to essentially eliminate
the model dependence in the MEs~\cite{slp2}.

\begin{figure}[tbp]
\epsffile{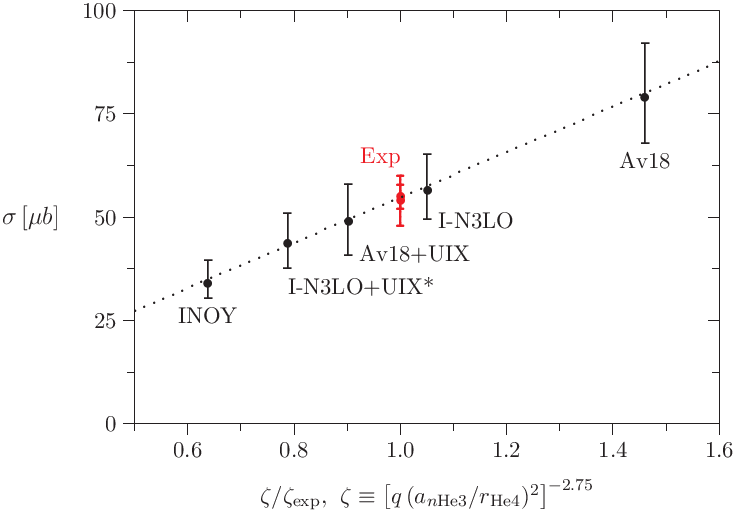}
\caption{\label{sigmacorr}
(Color online)
The $hen$ cross section $\sigma$ (in units of $\mu b$)
plotted against $\zeta/\zeta_{\rm exp}$,
where $\zeta= \left[q\,  (a_{n\rm{He3}}/r_{\rm{He4}})^2 \right]^{-2.75}$,
and $\zeta_{\rm exp}$ is the value of $\zeta$
when all the quantities therein
are given their respective empirical values
(See Table \ref{tab:be_sigma_hen}).
For each indicated nuclear interaction,
the error bar represents the range of variation
for three different choices of the cutoff parameter:
$\Lambda=500$ MeV (lower end),
700 MeV (filled circle) and 900 MeV (upper end).
The lines in red with label ``Exp" denote the experimental data,
$54\pm 6\ \mu b$ \cite{Wolfs} and $55\pm 3\ \mu b$  \cite{Wervelman}.}
\end{figure}

We expect that in principle a similar procedure can be adopted
for the $hen$ process.
To this end, it seems useful to find a quantity
that is related to the ERS in the $A$=4 systems
and that exhibits strong correlation with the $hen$ cross section,
$\sigma$.
We define the quantity $\zeta$ by
\be
\zeta \equiv \left[q\,  (a_{n\rm{He3}}/r_{\rm{He4}})^2 \right]^{-2.75}\,,
\label{zeta}
\ee
where $q={\rm BE}(\He4)-{\rm BE}(\He3)$, 
and the other quantities have already appeared in Table I.
Fig.~\ref{sigmacorr}  shows the calculated  values 
of $\sigma$ as a function of $\zeta$.
Strong linear dependence
between $\sigma$ and $\zeta$ can be seen.
Suppose we take this correlation seriously and consider
the quantity $\tilde \sigma$ defined by
\be
\tilde\sigma \equiv \frac{\zeta_{\rm exp}}{\zeta}
\,  \sigma\,,
\ee
where $\zeta_{\rm exp}$ is the value of $\zeta$
when all the quantities in eq.(\ref{zeta})
are given their respective empirical values.
Then $\tilde \sigma$ turns out to be
almost model-independent:
\be
\tilde \sigma = ( 54.8\pm 8.4,\ 54.5\pm 7.5,
\ 53.9\pm 7.1,\ 54.8\pm 9.4,\ 56.4\pm 8.5)\ \mu b
\label{sigma-hen-corr}\ee
for AV18, I-N3LO, INOY, AV18+UIX and I-N3LO+UIX*,
respectively,
and these values are all in agreement with data.

It is however not quite clear whether
the correlation between $\sigma$ and $\zeta$
is accidental or physical in nature.
Furthermore, since $hen$ is a four-body process
that involve a large number of ERPs to be controlled,
our numerical results that only cover
five different nuclear interactions
may not be sufficient to establish the meaning of
correlation unambiguously.
For example,
a correlation similar to the one between $\sigma$-$\zeta$
can be seen if we plot $\sigma$ against
$q^5 P_D^{2/3}$.
(It is however not clear whether this
correlation is independent of
the $\sigma$-$\zeta$ correlation.
They can be just different ways
to express the same correlation,
because $P_D$ and  $r_{\rm{He4}}$
may not be independent of each other.)
Without delving into the discussion of physics
behind the $\sigma$-$\zeta$ correlation,
we take here the viewpoint  that
the $\sigma$ calculated with a nuclear interaction
that does not reproduce the relevant ERPs
should be considered much less reliable
than the $\sigma$ obtained with a nuclear interaction
that does reproduce the relevant ERPs.
Based on this viewpoint,
we adopt here the results obtained with AV18+UIX and I-N3LO+UIX*
as the most reliable theoretical values for $\sigma$.
For these two cases,  the calculated $hen$ cross sections are:
\be
\sigma= 49.4 \pm 8.5 \ \mu b \,\,{\rm (AV18+UIX)},
\,\,\,\,\,\,\,\,\sigma= 44.4 \pm 6.7\ \mu b\,\,
{\rm (I}\!-\!{\rm N3LO}\!+\!{\rm UIX}^*{\rm )}.
\label{sigma-hen}
\ee
These values are in good agreement with the data
within experimental errors.

\subsection{Comparison of our results with previous studies\label{comparison}}

Making comparison of our calculation with the other related studies
is not straightforward  
due to differences in the employed current operators
and the nuclear wave functions;
%
here we limit ourselves to discussion of the 1B:LO contribution.
Let $\sigma_{\rm 1B:LO}$ be the $hen$ cross section
arising from the 1B:LO term.
In our calculation,
$\sigma_{\rm 1B:LO}$ varies from 4 $\mu$b (for INOY) to
$\sim$16 $\mu$b (for AV18),
which agrees well with the range  of values
$(2\sim 14)\ \mu$b obtained by Towner and
Khanna~\cite{Towner:1981hz} using rather schematic wave functions.

It seems more significant to compare our results
with those of Ref.~\cite{Carlson:1990nh},
where the authors used the variational Monte-Carlo (VMC) method
with the AV14+UVII potentials
and obtained $\sigma_{\rm 1B:LO}=5.65\ \mu$b.
%
Our calculation with AV18+UIX gives
$\sigma_{\rm 1B:LO}=14.9\ \mu$b,
which is about 2.6 times larger than the result of
Ref.~\cite{Carlson:1990nh}.
Since the 1B:LO M1 operator is the same for both cases, the discrepancy should be traced to
the difference in the nuclear wave functions.
Part of the discrepancy can be attributed
to the fact that the calculated values of the ERPs
vary for different realistic nuclear interactions used.
For the $n$-$\He3$ scattering length $a_{n\rm{He3}}$,
Ref.\cite{Carlson:1990nh} reports
$\Re(a_{n\rm{He3}})=3.5\ \fm$ for AV14+UVII,
and that $\sigma_{\rm 1B:LO}$ is increased by about 40 \%
if $a_{n\rm{He3}}$ is reduced from 3.5 fm to 3.25 fm,
which is close to $3.23\ \fm$ for AV18+UIX.
However this feature can explain
only small part of the discrepancy.

An explanation for the remaining discrepancy
may lie in the fact that the large-$r$ boundary condition
for the initial scattering state
adopted in Ref.~\cite{Carlson:1990nh}
does not take into account coupling
between the $n$-$\He3$ and the $p$-$\H3$ states.
Such a simplification may be considered
rather unwarranted for the following reasons.
First,
the presence of the diverging factor $\sqrt{p_p/p_n}$
in eq.(\ref{eq_as_beh})
enhances the coupling in the asymptotic region.
%
Secondly, ignoring the coupling
between $n$-$\He3$ and $p$-$\H3$ states
breaks orthogonality
between the incoming and outgoing wave functions.
Since, as mentioned, the 1B:LO contribution undergoes
a huge suppression due to the pseudo-orthogonality,
even a small breaking of the orthogonality
can have very strong influence on the 1B:LO contribution.
%
%
%
%
Finally, the variational calculation in~\cite{Schiavilla:1992sb}
involves too few correlation operators
to accurately describe
the non S-wave components
(the ones coupled by the IA operator) in the variational
wave function.

It is noteworthy that
a very similar feature occurs in $hep$ calculations.
That is, the correlated-hyperspherical harmonics
method~\cite{Marcucci:2000bh}
with AV18+UIX leads to a 1B $S$-factor that is
about four times larger than that obtained in the VMC
calculation~\cite{Schiavilla:1992sb} with the AV14+UVIII;
this additional example seems to render support
to our above argument.

As mentioned in section \ref{sec:intro},
after the submission of this work, 
there has appeared an elaborate calculation of $hen$ \cite{Girlanda:2010vm},
in which both the potential and current operators 
are derived from chiral effective field theory
using time ordered perturbation theory.
Ref. \cite{Girlanda:2010vm} gives
$50\pm 6\ \mu b$ for the total $hen$ cross section.
Despite the differences in the details of the formulation, 
the general features of the calculation and the numerical results
are in agreement between our work and Ref. \cite{Girlanda:2010vm}.

\subsection{Cutoff dependence}

We now turn our attention to the cutoff dependence.
Table~\ref{tab:be_M_hen_cutoff} shows to what extent
the $hen$ matrix element ${\cal{M}}$
calculated for the AV18+UIX wave functions changes
when the cutoff parameter $\Lambda$
is varied over a range $\Lambda=500\sim900$ MeV.
\begin{table}[btp]
\caption{Cutoff-dependence of the $hen$ matrix element ${\cal{M}}$
calculated with the AV18+UIX wave functions.
The $hen$ cross section $\sigma$ is in units of $\mu b$;
for other explanations,
see the caption for Table~\ref{tab:be_M_hen700}.
}
\label{tab:be_M_hen_cutoff}%
\begin{ruledtabular}
\begin{tabular}{crrccc}
$\Lambda$ [MeV] & $g_{4s}$ & $g_{4v}$ & $\Re \calM$ & $\Im \calM$ &
$\sigma$\\ \hline
500 & 0.8366 & 1.9068 & $-0.0915 + 0.0246\, g_{4s} -0.0471\, g_{4v}
= -0.1608$ & 0.0166 & 40.9 \\
600 & 0.6990 & 0.5886 & $-0.1593 + 0.0231\, g_{4s} -0.0433\, g_{4v}
=-0.1686$ & 0.0173 & 44.9 \\
700 & 0.5810 & $-0.4615$ & $ -0.2049 + 0.0205\, g_{4s} -0.0377\, g_{4v}
= -0.1756$ & 0.0179 & 48.7 \\
800 & 0.4517 & $-1.3622$ & $ -0.2346 + 0.0176\, g_{4s} -0.0319\, g_{4v}
= -0.1832 $ & 0.0186 & 52.9 \\
900 & 0.3169 & $-2.2069$ & $-0.2547 + 0.0149\, g_{4s} -0.0265\, g_{4v}
= -0.1915$ & 0.0195 & 57.9
\end{tabular}
\end{ruledtabular}
\end{table}
Table~\ref{tab:be_M_hen_cutoff} indicates that
the renormalization procedure of the LECs,
$g_{4s}$ and $g_{4v}$,
plays an essential role in
reducing the cutoff-dependence.
As a way of quantifying this feature,
we define the quantity
\be
R\equiv \frac{\calM_{\rm total}(\Lambda_2) - \calM_{\rm total}(\Lambda_1) }{
 \calM_{\rm finite}(\Lambda_2) - \calM_{\rm finite}(\Lambda_1)},
\ee
where the subscript ``finite" stands for ``finite-range term contributions",
and $\calM_{\rm finite}$ corresponds
to a case where all the terms other than the
contact term (CT) contributions are included.
Perfect renormalization invariance would correspond to $R=0$.
It turns out that
$R_{hen}=0.189$ for $(\Lambda_1,\,\Lambda_2)=(500,\,900)$ MeV.
Thus the renormalization procedure of LECs has removed
a major part of cutoff-dependence;
the cutoff-dependence of $\calM_{\rm total}$
is about one-fifth of that of $\calM_{\rm finite}$.
It is interesting to compare the above value of $R_{hen}$
with the corresponding quantity $R_{hep}$
obtained in a $hep$ calculation in~\cite{hep}.
The $hep$ calculation in \cite{hep}
is based on the same MEEFT strategy
(but uses a different method for obtaining
exact solutions to the nuclear Schr\"{oe}dinger equations).
Ref.~\cite{hep} reports $R_{hep}=0.137$
for the slightly smaller window,
$(\Lambda_1,\,\Lambda_2)=(500,\,800)$ MeV.
Thus the previous $hep$ calculation~\cite{hep} is consistent
with the $hen$ calculation in the present work,
and this consistency provides further support
to the $hep$ results in \cite{hep}.

\subsection{Convergence of chiral expansion}

\begin{table}[btp]
\caption{The magnetic moments, $\mu(\H3)$ and  $\mu(\He3)$,
and the real and imaginary parts of the $hen$ matrix element,
$\calM$,
calculated for the AV18+UIX wave functions
and for $\Lambda=700$ MeV.
The values of $\mu(\H3)$ and $\mu(\He3)$ given in the row
labeled ``Total" are experimental values, 
which are used to fix the LECs.
The LECs corresponding to this case are:
$(g_{4s},\, g_{4v})=(0.581,\, -0.4615)\, [\fm^3]$.
}
\vspace{3mm}
\label{tab:details}%
\begin{ruledtabular}
\begin{tabular}{lrrrr}
 & $\mu(\H3)$ & $\mu(\He3)$ & $\Re \calM$ & $\Im \calM$ \\ \hline
\hline
1B: LO  & $2.5727$ & $-1.7632$ & $0.0964$ & $-0.0136$ \\ 
1B: RC  & $-0.0171$ & $0.0037$ & $0.0554$ & $-0.0075$ \\ 
1B-total	& $2.5556$ & $-1.7595$ & 0.1518 & $-0.0211$ \\ \hline
2B: ${1\pi}$ (NLO)	    & 0.2292 & $-0.2258$ & $-0.1657$ & 0.0195\\	
2B: ${1\pi}C$ (\nlo3)	& 0.1578 & $-0.1289$ & $-0.1465$ & 0.0172 \\	
2B: ${2\pi}$	(\nlo3)	& 0.0419 & $-0.0408$ & $-0.0445$ & 0.0052\\ \hline
finite (total w/o CT) &2.9845& $-2.1550$ & $-0.2049$  & 0.0208	\\ \hline
	& $0.0193\, g_{4s}$ & $0.0190\, g_{4s}$ & $0.0205\, g_{4s}$ & $-0.0014\, g_{4s}$ \\
	& $+ 0.0363\, g_{4v}$ & $-0.0354\,g_{4v}$ & $-0.0377\,g_{4v}$ & $+ 0.0044\,g_{4v}$ \\
2B: CT (\nlo3)   & $= -0.0055$ & $= 0.0274$ & $= 0.0293$ & $= -0.0029$ \\ \hline
Total		&2.9790& $-2.1276$ & $-0.1756$ &	0.0179	
\end{tabular}
\end{ruledtabular}
\end{table}

Table~\ref{tab:details} shows the individual contributions
of the various 1B and 2B terms to
$\mu(\H3)$, $\mu(\He3)$ and the $hen$ matrix element $\calM$,
calculated at $\Lambda=700$ MeV for the AV18+UIX potential.
We can see that the 1B contribution to $hen$ is highly suppressed
due to the aforementioned orthogonality
between the initial and final wave functions.
The NLO contribution, which comes from the {\em soft} one-pion-exchange,
is also suppressed for the M1 channel, due to the accidental cancelation between
the pion-pole and pion-seagull diagram contributions~\cite{slp2}.
These suppression mechanisms make chiral convergence rather unclear.
For example, one might worry about the fact
that the ${1\pi}C$ and ${2\pi}$ contributions,
both of which are \nlo3, turn out to be comparable in size
to the NLO ${1\pi}$ contribution.
It should be noted, however,
that most of the ${1\pi}C$ and ${2\pi}$ contributions
are to be absorbed in the renormalization of the LECs,
leaving very small net effects on the observable quantities.
To demonstrate this point,
we define the {\em effective} matrix element,
$\langle {\cal O}\rangle^{\rm effective}$,
of a given operator ${\cal O}$ by
\be
\langle {\cal O}\rangle^{\rm effective} \equiv \calM_{total} -
\left(\calM_{total}\ \mbox{but without}\ \langle{\cal O}\rangle\right).
\ee
Thus $\langle {\cal O}\rangle^{\rm effective}$
represents a net change in the amplitude
that would occur if we omit the operator ${\cal O}$.
In evaluating the parenthesized quantity,
the LECs should be readjusted so as to reproduce the experimental values
of the $A=3$ magnetic moments
{\em without} $\langle {\cal O}\rangle$;
because of this readjustment we should expect
$\langle {\cal O}\rangle^{\rm effective}\ne
\langle {\cal O}\rangle$.
We find
\be
\langle {1\pi}\rangle^{\rm effective} &=& \ \  0.0749 -0.0087i,
\ \ \ (g_{4s}, g_{4v})= (0.5223,\, 5.8848)\ \fm^3,
\nonumber \\
\langle {1\pi}C\rangle^{\rm effective} &=& -0.0093+ 0.0004i,
\ \ \ (g_{4s}, g_{4v})= (1.2433,\, 3.53455)\ \fm^3,
\nonumber \\
\langle {2\pi}\rangle^{\rm effective} &=& -0.0010+0.0001i,
\ \ \ (g_{4s}, g_{4v})= (0.5843,\, 0.6921)\ \fm^3,
\label{effME}\ee
where we have also listed the corresponding values of the LECs,
which should be compared with $(g_{4s}, g_{4v})$=$(0.581,\,-0.4615)\,\fm^3$
that correspond to the full calculation up to \nlo3.
Eq.(\ref{effME}) demonstrates that the {\it effective} contributions
of ${1\pi}C$ and ${2\pi}$ are very small,
only about 6 \% and 2 \%,  respectively,
relative to the values one would naively expect.
This is in sharp contrast with the NLO soft one-pion-exchange,
whose contribution cannot be absorbed in the LECs.
A rigorous examination of chiral convergence would require
a calculation that goes one order higher than considered in the present work
({\it i.e.,} we need to go up to \nlo4), but we relegate
this task to future studies.

There can also be a fully consistent EFT approach
where both nuclear interactions
and transition operators are obtained in the same EFT framework.
This approach requires much more involved calculations than MEEFT,
but recent significant progress in constructing EFT Hamiltonians
makes it more attractive.
Also available is a pionless EFT approach~\cite{pionless}
where the matrix elements are evaluated perturbatively,
but, unless it is capable of reproducing all the
relevant ERPs of the nuclear systems involved,
its usefulness is limited.

\section{Discussion and Conclusions}

In this work we have performed
an {\it ab initio} parameter-free calculation
for the $hen$ cross section $\sigma$,
with the use of the EM currents
that have been derived from HBChPT up to \nlo3.
The exact nuclear wave functions for the initial and
final states have been obtained
by solving the Faddeev-Yakubovsky equations
for realistic nuclear interactions.
The calculated value of $\sigma$
shows a high degree of stability
as the cutoff parameter $\Lambda$ is varied
over a wide range, $\Lambda=(500\sim900)$ MeV,
and we obtain as the best estimate
$\sigma= 49.4\pm 8.5\ \mu b$ for AV18+UIX
and $44.4\pm6.7\ \mu b$ for I-N3LO+UIX*.
These values are in good agreement with the data,
$54\pm 6\ \mu b$~\cite{Wolfs} and $55\pm 3\ \mu b$~\cite{Wervelman}.

The successful application of MEEFT to $hen$
renders strong support to the previous MEEFT calculation of $hep$
in Ref.~\cite{hep};
furthermore, it demonstrates the basic soundness of
the MEEFT approach in general.
The present treatment is open to several improvements
such as: the inclusion of the next order terms
in chiral perturbation, in particular
the incorporation of the three-nucleon currents;
a more stringent control of mismatch in the
chiral counting between SNPA and a formally accurate chiral expansion that
enters in the currents; a better understanding of the role
the counter terms play in the renormalization group property.
It is reasonable, however, to expect
that the effects of these improvements
are essentially accommodated in the above-quoted error estimate
based on the cutoff dependence.
A robust estimation of the $hep$ S-factor
has been a long-standing challenge
in nuclear physics~\cite{Bahcall:2000nu}.
We believe that our MEEFT calculations of $hep$ and $hen$
have solved this problem to a satisfactory degree.

\section*{Acknowledgement}

The work of TSP is supported by the
Korea Science and Engineering Foundation (KOSEF) Basic Research Program with
the grant No. R01-2006-10912-0,
by the KOSEF grant funded by the Korea Government (MEST)
(No. M20608520001-08B0852-00110)
and
by the US National Science Foundation, Grant No. PHY-0758114.

YHS would like to express his gratitude to
Professor Dong-Pil Min who encouraged him to start this work.
The work of YHS was partly supported by the Korea Research Foundation Grant
funded by the Korean Government (MOEHRD, Basic Research Promotion Fund)
(KRF-2008-357-C00021) and the US Department of Energy 
under Contract No. DE-FG02-09ER41621.

We are deeply obliged to
Professors Mannque Rho and Kuniharu Kubodera for kindly agreeing to read the
manuscript prior to its publication and giving us important comments.

The numerical calculations have been
performed at IDRIS (CNRS, France).
We thank the staff members of the IDRIS
computer center for their constant help.

\end{document}